\documentclass[12pt,A4paper]{article}

\usepackage{amsmath}
\usepackage{epsfig,amssymb}
\usepackage{graphicx,psfrag}

\usepackage{color}
\begin{document}

\title{\bf Advanced memory effects in the aging of a polymer glass}

\author{L. Bellon, S. Ciliberto and C. Laroche \\ \'Ecole
Normale Sup\'erieure de Lyon, Laboratoire de Physique ,\\ C.N.R.S. UMR5672, \\
46, All\'ee d'Italie, 69364 Lyon Cedex 07, France\\}

\maketitle

\begin{abstract}
A new kind of memory effect on low frequency  dielectric
measurements on plexiglass (PMMA) is described. These measurements
show that cooling and heating the sample at constant rate give an
hysteretic dependence on temperature of the dielectric constant
$\epsilon$. A temporary stop of cooling produces a downward
relaxation of $\epsilon$. Two main features are observed i) when
cooling is resumed $\epsilon$ goes back to the values obtained
without the cooling stop ( i.e. the low temperature state is
independent of the cooling history) ii) upon reheating $\epsilon$
keeps the memory of all the cooling stops({\it Advanced memory}).
 The dependence of this effect on frequency and on the cooling
rate is analyzed. The memory deletion is studied too. Finally the
results  are compared with those of similar experiments done in
spin glasses and with the famous experiments of Kovacs.

\end{abstract}

\medskip
{\bf PACS:} 75.10.Nr, 77.22Gm, 64.70Pf, 05.20$-$y.


\newpage

\section{introduction}

Glasses are out of equilibrium systems, which are characterized by
a slow relaxation, which is often much longer than any laboratory
scale. One of the main consequences is that after a quench into
the glassy phase the properties of a glassy material depend on the
time spent below the glass transition temperature $T_g$, that is
to  say  the system ages.

 The aging of glassy materials is a widely studied phenomenon
\cite{Struick,book}.  In spite of the interesting experimental
\cite{Vincent,Alberici,Nagel,VincentPRL,Jonsson1,Doussineau,vigier,Bellon,Europhys,Jonason2000,Dupuis}
and theoretical progress
\cite{book,Bouchaud,Mezard,Bouchaud1,Kurchan}, done in the last
years, the physical mechanisms of aging  are not yet fully
understood. In fact on  the basis of  available experimental data
it is very difficult to distinguish which  is the most suitable
theoretical approach for describing  the aging processes of
different materials. In order to give more insight into this
problem several experimental procedures have been proposed and
applied to the study of the aging of various materials, such
 as spin-glasses (SG)\cite{Vincent,VincentPRL,Jonsson1,Jonason2000,Dupuis}, orientational
glasses (OG)\cite{Alberici,Doussineau}, polymers
\cite{Struick,vigier,Bellon,Europhys,Kovacs} and supercooled
liquids (SL)\cite{Nagel}.

The very well known Kovacs\cite{Kovacs} experiment was probably
among the first in showing that the state of a glass is strongly
determined by its thermal history. In the Kovacs experiment the
material is submitted to the following temperature cycle. The
material is rapidly quenched from the liquid to the glass phase at
a temperature $T_1$ at which the volume of the glass is $V_1$.
Because of aging the volume of the material decreases as a
function of time going from $V_1$ to $V_2$. Then the material is
rapidly heated  at a temperature $T_2$ at which $V_{2}$ is the
equilibrium volume. One could think that no appreciable evolution
appears as a function of time. Instead the volume first increases,
reaches a maximum and then decreases to reach again the value
$V_2$. The amplitude of the maximum depends of course on $T_1$.
Another example of history dependent effect is that induced by the
 aging at low temperature, known as prepeak in polymer
 literature \cite{vigier}. Such an aging  does not
produce important changes in the enthalpy. However when increasing
temperature, a peak in the curve of the heat capacity versus
temperature  is observed at a temperature close to $T_g$.

Other procedures were proposed for spin glasses and applied to
other materials.  Among these procedures we may recall the
applications of small temperature cycles to a sample during the
aging time\cite{Vincent,Alberici,Nagel,Bellon}. These experiments
have shown three main results in different materials: i)  there is
an important difference between positive and negative cycles and
the details of the response to these perturbations are material
dependent \cite{Vincent,Alberici,Bellon}; ii) for SG
\cite{Vincent} the time spent at the higher temperature sort of
reinitialize the aging at a lower temperature whereas for
plexiglass (PMMA) \cite{Bellon} and OG \cite{Alberici} it only
slightly modifies the long time behavior; iii) A memory effect has
been observed for negative cycles. Specifically when temperature
goes back to the high temperature the system recovers its state
before perturbation. In other words the time spent at low enough
temperature does not contribute to the aging behavior at the
higher temperature.

In order to have a better understanding of the free energy
landscape of SG and OG, a new cooling protocol has been proposed
\cite{VincentPRL} and used in several experiments
\cite{VincentPRL,Jonsson1,Doussineau}. This protocol, which is
characterized by a temporary cooling stop, has revealed that in SG
and in  OG  the low temperature state is independent of the
complete cooling history but that these materials keep the memory
of all the aging history ({\it Memory effect}) \cite{VincentPRL}.

In a recent paper \cite{Europhys} we described an  experiment
where we have used the cooling protocol, proposed in
ref.\cite{VincentPRL}, to show that
 memory  effects are present during the aging of the
dielectric constant of plexiglass (PMMA), which is a polymer glass
with $T_g=388K$
 \cite{bookp}.
The purpose of this article is to extend the results of
ref.\cite{Europhys} and to show that more complex temperature
cycle can be observed. The deletion effect is also discussed.
Finally  we  compare the behavior of PMMA to that of SG and OG,
submitted to the same cooling protocol.

As we already mentioned several kinds \cite{vigier,Kovacs} of
"memory effects" for polymers   have been described in literature.
The most famous one  is that of Kovacs \cite{Kovacs}. However the
thermal cycles used in these experiments  are quite different from
those proposed in \cite{VincentPRL}. Thus it is interesting to
check how polymers behave when they are submitted to this new
thermal procedure, which can give new insight on the aging of
these materials. Moreover, the use of the same protocols makes the
comparison between different kinds of glasses easier.

The paper is organized as follows. In section 2  we describe the
experimental set-up and we discuss the results of a single quench
experiment. The aging properties of the PMMA dielectric constant
are studied as a function of the quench temperature and of the
measuring frequency. In section 3 the measure and the features of
the memory effect are analyzed.  In section 4 the memory deletion
and the double memory effect are described. Finally in section 5
the results on the PMMA dielectric constant are discussed and
compared with those observed in other systems.

\section{Aging range: a simple quench experiment}

\begin{figure}[!ht]
    \begin{center}
        \psfrag{E}[bl][bl]{\tiny oven}
        \psfrag{P}[cl][cl]{\tiny PMMA ($\text{\o} = 10cm$, $e = 0,3mm$)}
        \psfrag{C}[tl][tl]{\tiny Capacitor (copper)}
        \psfrag{T}[cl][cl]{\tiny $\textrm{Pt}_1$}
        \psfrag{Y}[cl][cl]{\tiny $\textrm{Pt}_2$}
        \psfrag{S}[cr][cr]{\tiny $\textrm{Pt}_2$}
        \psfrag{D}[cl][cl]{\tiny $Z$}
        \psfrag{B}[Bc][Bc]{\tiny electric}
        \psfrag{F}[Bc][Bc]{\tiny connections}
        \includegraphics[width=7cm]{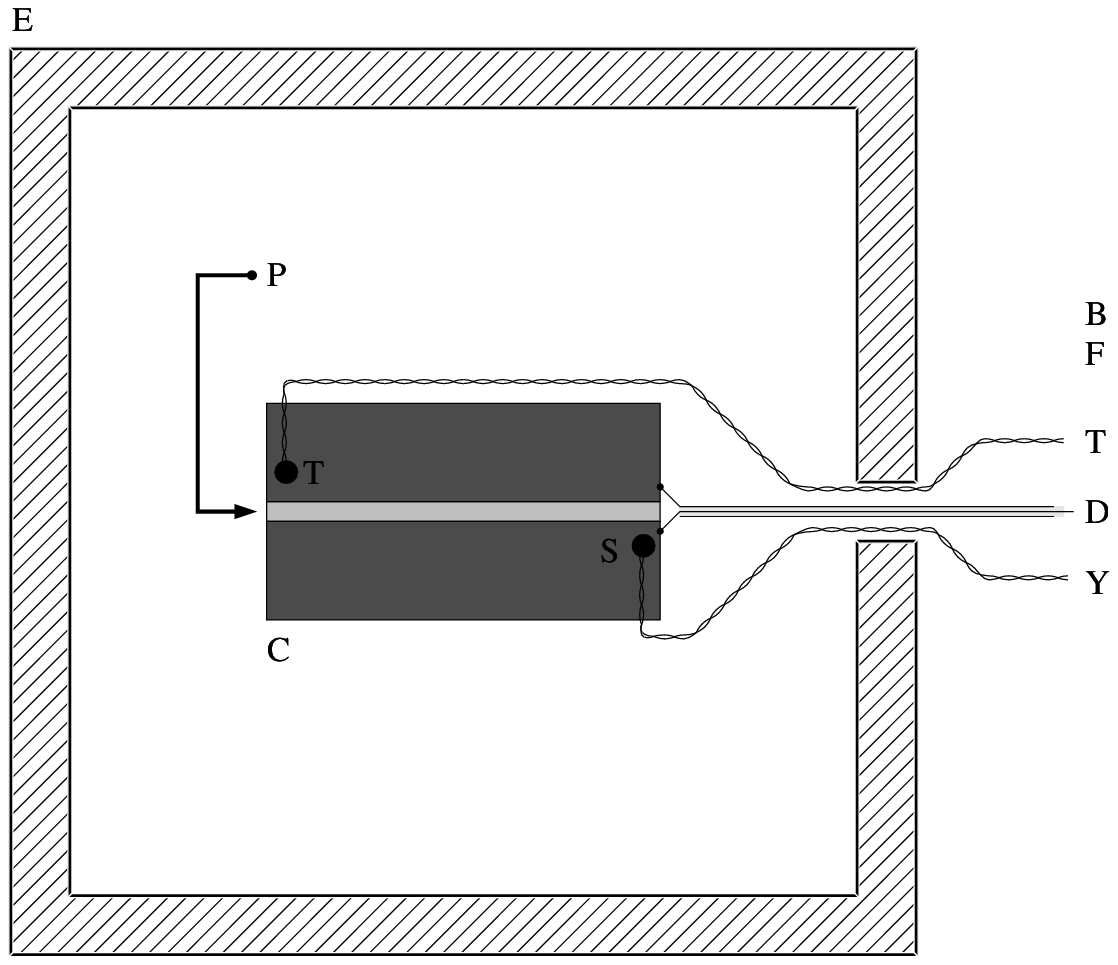}
    \end{center}
  \caption{Experimental set-up}{PMMA is the dielectric of a capacitor whose vacuum capacitance
  is  $C_0=230pF$. This impedance   $Z$ is inside an oven which controls the temperature and
  acts like an electrical screen.  The dielectric temperature is measured by two
  platinum probes (Pt)}
  \label{fig:experiment}
\end{figure}

To determine the dielectric constant, we measure the complex
impedance of a capacitor whose dielectric is the PMMA sample. In
our experiment a disk of PMMA of diameter $10cm$ and thickness
$0.3mm$ is inserted between the plates of a capacitor whose vacuum
capacitance is $C_o=230 pF$ (see fig.\ref{fig:experiment} for
details). The capacitor is inside an oven whose temperature $T$
may be changed from $300 K$ to $500 K$. The temperature stability
is  within $0.1 K$.   The maximum heating and cooling rate
$|R|=|dT/dt|$ is about $180 K/h$. We checked that the temperature
difference between the two capacitor plates is always smaller than
$1K$ both during the heating and the cooling cycles.

The capacitor is a component of the feedback loop of a precision
voltage amplifier whose input is connected to a signal generator.
We obtain the real and imaginary part of the capacitor impedance
by measuring the response of the amplifier to a sinusoidal input
signal. This apparatus allows us to measure the real and imaginary
part of the dielectric constant $\epsilon=\epsilon_1 +i \
\epsilon_2$ as a function of temperature $T$, frequency $\nu$ and
time $t$. Relative variations of $\epsilon$ smaller than $
10^{-3}$ can be measured in all the frequency range used in this
experiment, i.e. $0.1Hz < \nu <100Hz$. The following discussion
will focus only on $\epsilon_1$ (thus we will omit subscript 1),
for which we have the best accuracy, but the behavior of
$\epsilon_2$ leads to the same conclusions.

\begin{figure}
    \begin{center}
         \psfrag{T120}[Bl][Bl]{\tiny $395K$}
         \psfrag{T115}[Bl][Bl]{\tiny $387K$}
         \psfrag{T110}[Bl][Bl]{\tiny $382K$}
         \psfrag{T100}[Bl][Bl]{\tiny $375K$}
         \psfrag{T90}[Bl][Bl]{\tiny $365K$}
         \psfrag{T80}[Bl][Bl]{\tiny $354K$}
         \psfrag{T70}[Bl][Bl]{\tiny $343K$}
         \psfrag{T60}[Bl][Bl]{\tiny $332K$}
         \psfrag{T40}[Bl][Bl]{\tiny $311K$}
         \psfrag{Tstop}[bl][bl]{\footnotesize $T_{stop}=$}
         \psfrag{f0.2}[Bl][Bl]{\tiny $0.2Hz$}
         \psfrag{f0.5}[Bl][Bl]{\tiny $0.5Hz$}
         \psfrag{f1}[Bl][Bl]{\tiny $1Hz$}
         \psfrag{f2}[Bl][Bl]{\tiny $2Hz$}
         \psfrag{f5}[Bl][Bl]{\tiny $5Hz$}
         \psfrag{f10}[Bl][Bl]{\tiny $10Hz$}
         \psfrag{f20}[Bl][Bl]{\tiny $20Hz$}
         \psfrag{f=}[Bl][Bl]{\scriptsize $\nu=$}
         \psfrag{2}[bl][bl]{ }
         \psfrag{xl}[ct][ct]{\small Time $t$ (h)}
         \psfrag{yl}[Bc][Bc]{\small dielectric constant $\epsilon$}
        {\includegraphics[width=6cm]{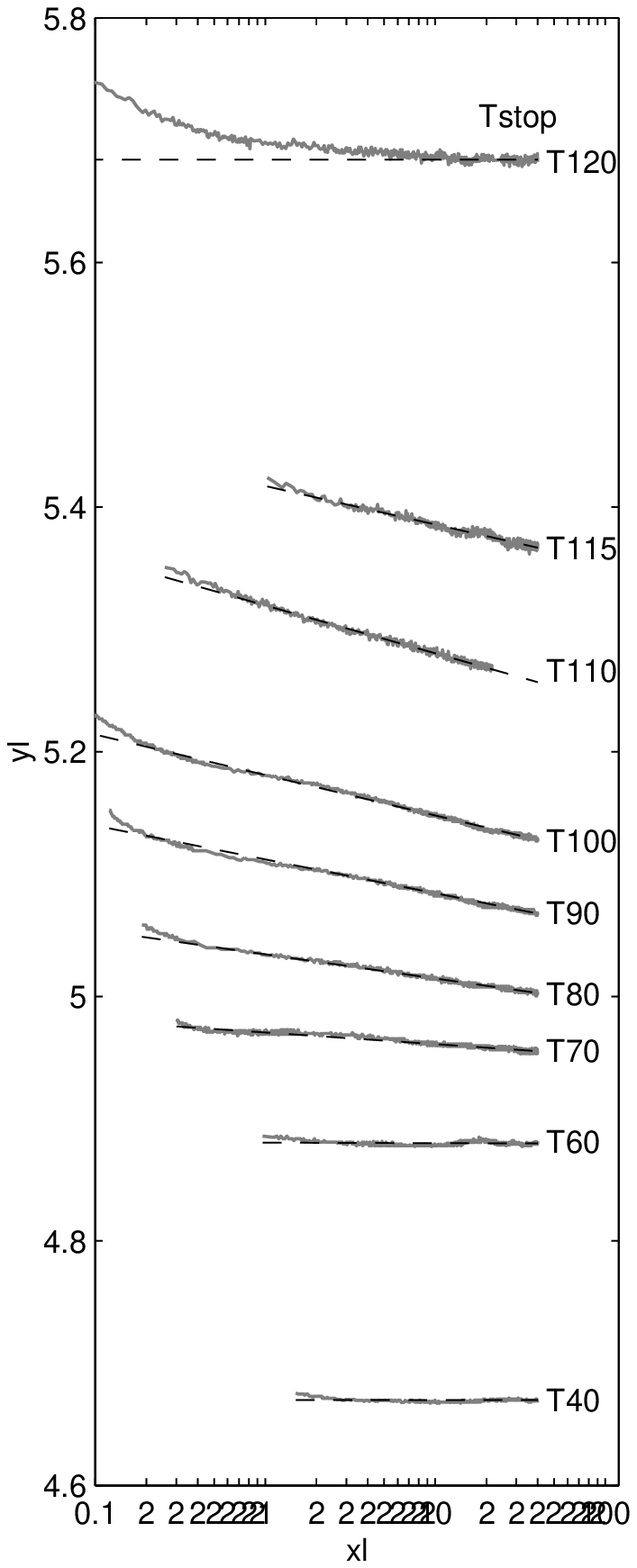}}
        \hspace{1mm}
    {\includegraphics[width=6cm]{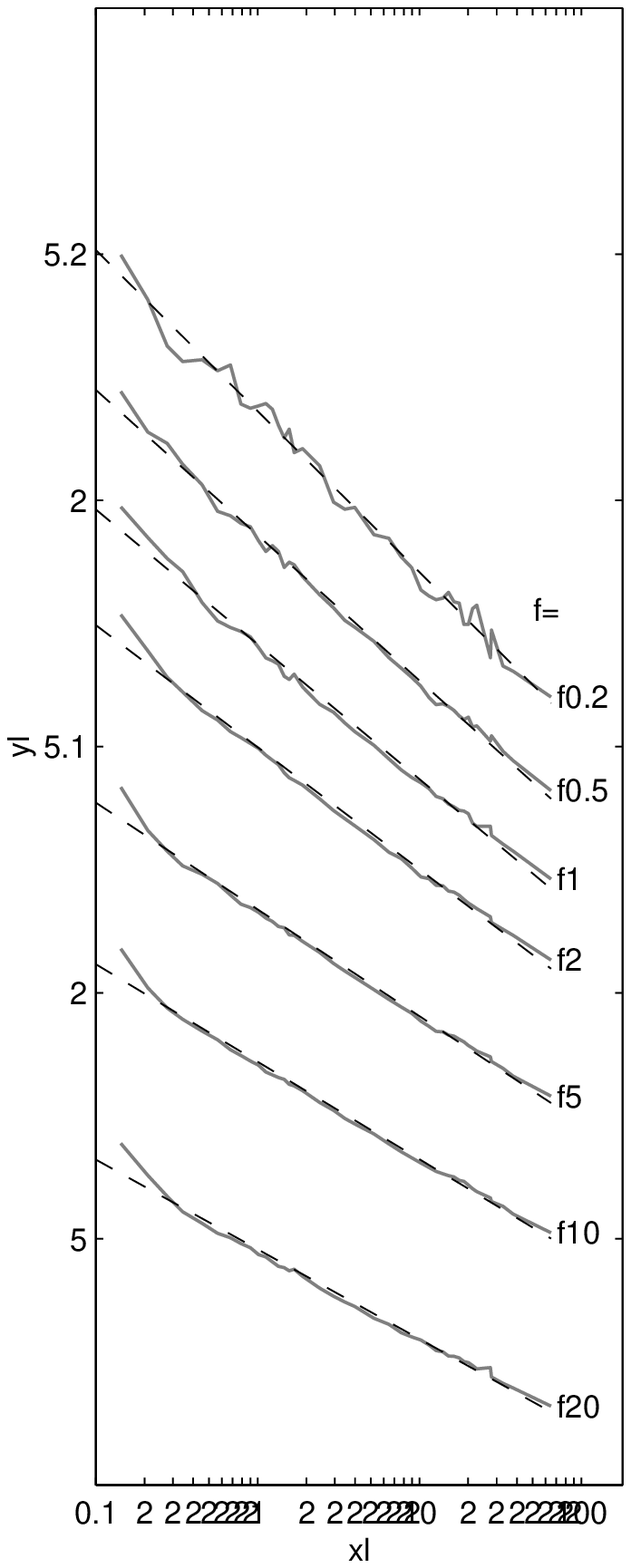}}
    \end{center}
\caption{Dependence on $t$ of $\epsilon$ after a quench. (a) Aging measured at $\nu=1Hz$ after a quench
at various $T_{stop}$. (b) Aging measured after a quench at
$T_{stop}=365K$ at various $\nu$. A logarithmic fit in time of all
these curves is accurate as soon as temperature is
stabilized, except for $T_{stop}=395K>T_g$} \label{fig:refagi}
\end{figure}

The measurement is performed in the following way. We first
initialize the PMMA history by heating the sample at a temperature
$T_{max}>T_g$. The sample is left at $T_{max}=415K$ for a few
hours, so that equilibrium can be assumed for the initial
condition. The temperature is rapidly decreased at $R=-180 K/h$ to
a temperature $T_{stop}<T_g$, where it is regulated by the oven.
The zero of the aging time is taken at the instant, during the
quench, when the sample temperature is equal to $T_g$. Typical
aging curves of $\epsilon$ as a function of time are shown in
fig.\ref{fig:refagi}, for different $T_{stop}$ at $\nu=1Hz$ and
different $\nu$ at $T_{stop}=365K$ . We clearly notice a
logarithmic dependence on time of the dielectric constant as soon
as the temperature is stabilized, in all the temperature and
frequency range we have explored. However the aging curve depends
on $T_{stop}$ and $\nu$ : the sample properties evolve faster when
$T_{stop}$ is close to $T_g$ and $\nu$ is small.

\begin{figure}
    \null
    \begin{center}
        {\psfrag{Tg}[cc][cc]{\colorbox{white}{\footnotesize $T_{g}$}}
            \psfrag{2}[bl][bl]{ }
            \psfrag{xl}[ct][ct]{\small Temperature $T_{stop}$ (K)}
            \includegraphics{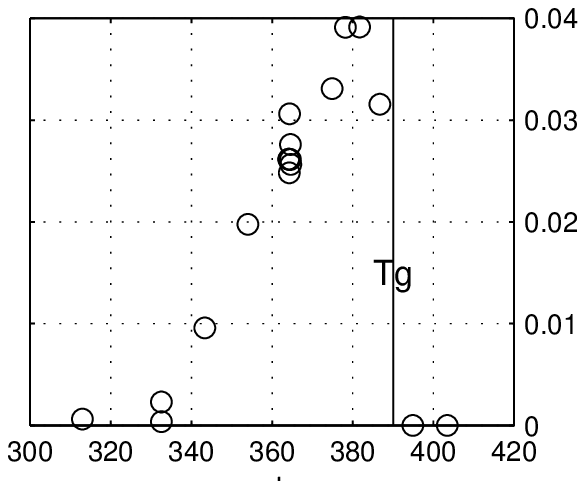}}
        \hspace{-1mm}
    {\psfrag{2}[bl][bl]{ }
            \psfrag{xl}[ct][ct]{\small Frequency $\nu$ (Hz)}
            \psfrag{yl}[Bc][Bc]{\small $B$}
            \includegraphics{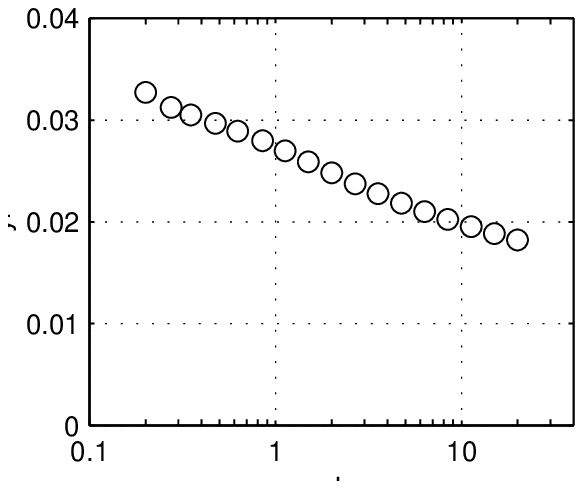}}
    \end{center}
\caption{(a) Dependence on $T_{stop}$ of $B$  at
$\nu=1Hz$, where $B$ is the parameter of the logarithmic fit of the
dielectric constant $\epsilon(T_{stop},t,\nu) = A(T_{stop},\nu) -
B(T_{stop},\nu) \ \log(t/1h)$ after a quench at $T_{stop}$. (b) Dependence on
$\nu$ of $B$ at $T_{stop}=365K$.}
\label{fig:alpha}
\end{figure}

Specifically one can write $\epsilon(T_{stop},t,\nu) =
A(T_{stop},\nu) - B(T_{stop},\nu) \ \log(t/t_o)$ with $t_o=1h$.
Here $A$ is the value of $\epsilon$ at $t=t_o$. It is found that
$A$ and $B$ are functions of $T_{stop}$ and of the frequency $\nu$
at which the dielectric constant is measured. Note that as A and B
depends also weakly on  the cooling rate $R$, we always use
$R=-180 K/h$ in these experiments.  The values of $B$, measured at
$\nu=1Hz$, are plotted in fig.\ref{fig:alpha}(a) as a function of
$T_{stop}$. $B$ is an increasing function of $T_{stop}$ till a
temperature $T_m$ close to $T_g$, but goes down to 0 if
$T_{stop}>T_g$. Indeed for a quench temperature larger than $T_g$
the sample can reach a thermodynamic equilibrium, so no aging is
observed but only a relaxation of $\epsilon$ toward its stationary
value. The long time behavior is stationary  and  $B$ goes to $0$.

The values of $B$, measured at $T_{stop}=365K$, are plotted in
fig.\ref{fig:alpha}(b) as a function of $\nu$. $B$ is a  slowly increasing function for
$\nu\rightarrow 0$. Indeed aging is smaller at high frequencies than at low
frequencies (a theoretical justification of such a behavior can be found for
example in ref.\cite{Bouchaud}).

These curves determine the region where to work. In our dielectric
measurement aging can be accurately observed between $335 K$ and
$T_g = 388 K$. We will probe $\nu = 0.1 Hz$ and $\nu = 1 Hz$ where
aging effects are the largest in our frequency range.

\section{Simple memory}

In this section, we describe experiments following the protocols
first proposed in ref.\cite{VincentPRL}: a temporary stop is done
during the cooling of the sample, and its consequence on the
dielectric constant behavior are studied. The measurement is
performed in the following way: again, the PMMA history is first
reinitialized by heating the sample at a temperature $T_{max}>T_g$
and leaving it at $T_{max}=415K$ for a few hours. Then it is
slowly cooled from $T_{max}$ to a temperature $T_{min}=313K$ at
the constant rate $R$ and heated back to $T_{max}$ at the same
$|R|$. The dependence of $\epsilon$ on $T$ obtained by cooling and
heating the sample at a constant $|R|$, is called the reference
curve $\epsilon_r$.

\begin{figure}
\begin{center}
\epsfig{figure=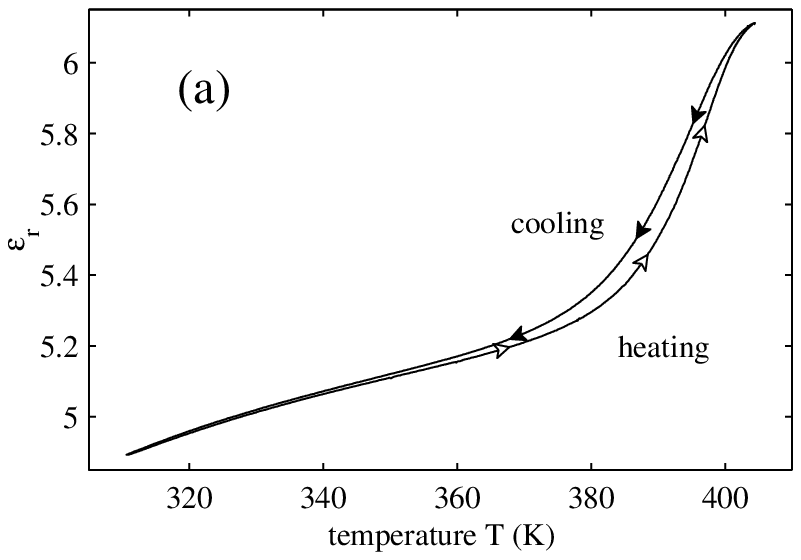} \epsfig{figure=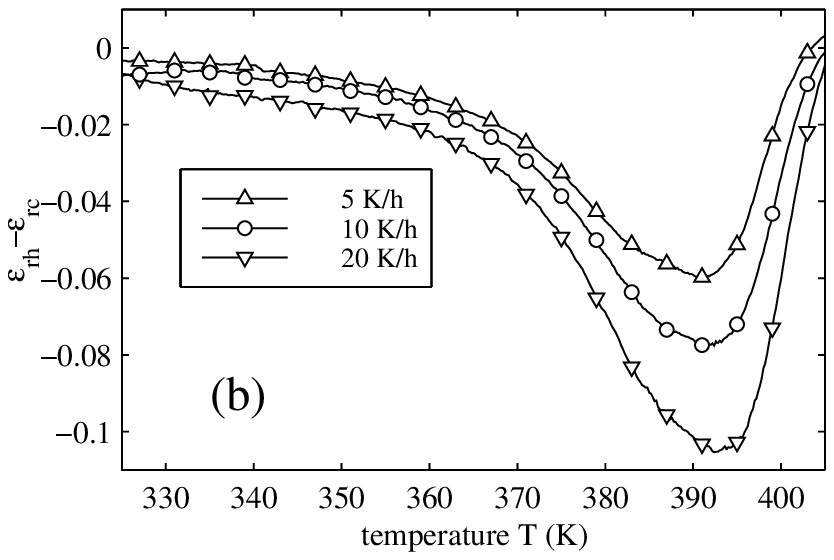}
\end{center}

\caption{(a) Evolution of $\epsilon_r$ at $\nu=0.1Hz$ as a function of $T$.
Reference curve obtained with $|R|=20K/h$. (b) Hysteresis of the reference
curve (difference between the heating and cooling curves $\epsilon_{rh} -
\epsilon_{rc}$) for 3 different $|R|$: $5 K/h$ ($\vartriangle$), $10 K/h$
($\circ$) and $20 K/h$ ($\triangledown$).} \label{fig:tempcyc1}
\end{figure}

As an example of reference curve we plot in fig.\ref{fig:tempcyc1}(a)
$\epsilon_r$, measured at $0.1Hz$ and at $|R|=20K/h$. We see that $\epsilon_r$
presents a hysteresis between the cooling and the heating in the interval
$350K<T<405K$. This hysteresis depends on the cooling and heating rates.
Indeed, in fig.\ref{fig:tempcyc1}(b), the difference between the heating curve
($\epsilon_{rh}$) and the cooling curve ($\epsilon_{rc}$) is plotted as a
function of $T$ for different $|R|$. The faster we change temperature, the
bigger hysteresis we get. Furthermore the temperature of the hysteresis
maximum is a few degrees above $T_g$, specifically at $T\approx 392K$. The
temperature of this maximum gets closer to $T_g$ when the rate is decreased.

\begin{figure}
\begin{center}
\epsfig{figure=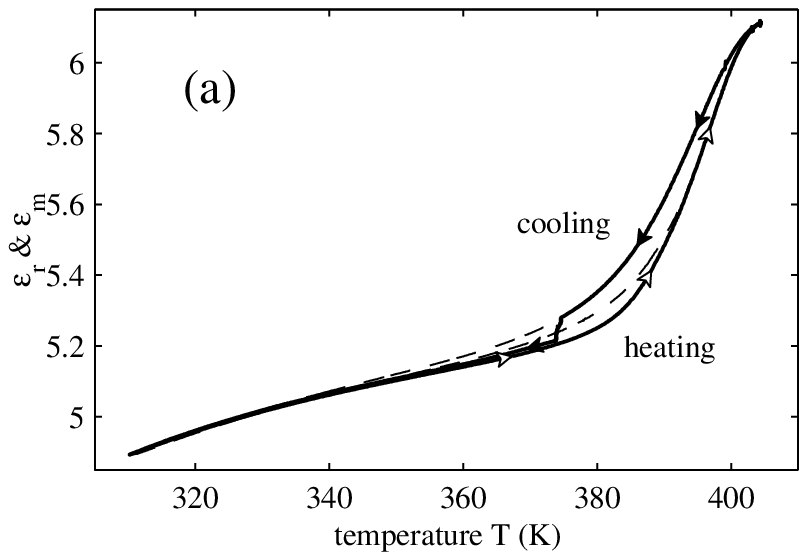} \epsfig{figure=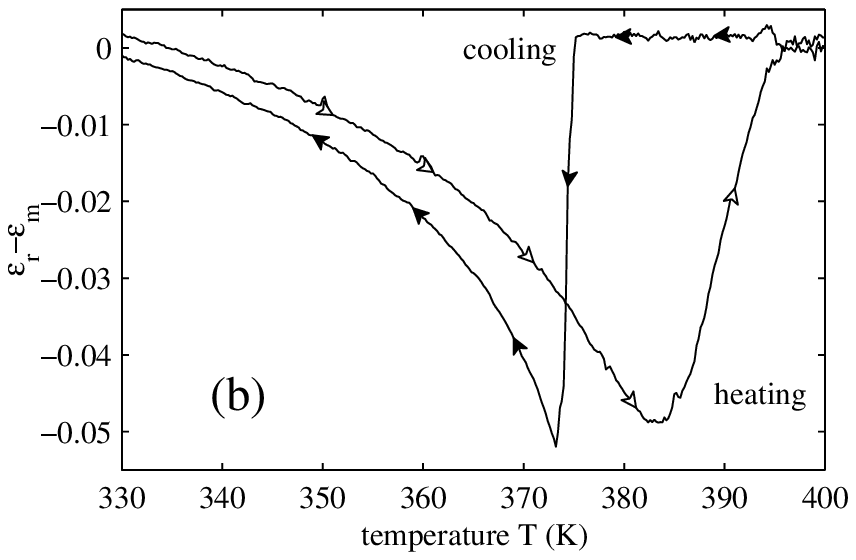}
\end{center}
\caption{(a)Evolution of $\epsilon$ at $\nu=0.1Hz$ as a function of $T$. The
dashed line corresponds to the reference curve ($\epsilon_r$) of
Fig.\ref{fig:tempcyc1}(a). The solid bold line corresponds to a different
cooling procedure: the sample is cooled, at $R=-20K/h$, from $T_{max}$ to
$T_{stop}=374K$, where cooling is stopped for $10h$. Afterwards the sample is
cooled at the same $R$ till $T_{min}$ and then heated again at $R=20K/h$ till
$T_{max}$. (b) Difference between the evolution of $\epsilon_r$ and
$\epsilon_m$. Downward filled arrows correspond to cooling
($\epsilon_{mc}-\epsilon_{rc}$) and upward empty arrows to heating
($\epsilon_{mh}-\epsilon_{rh}$).} \label{fig:tempcyc2}
\end{figure}

We neglect for the moment the rate dependence of the  hysteresis
and we consider as reference curve the one, plotted in
fig.\ref{fig:tempcyc1}(a), which has been obtained at $\nu=0.1Hz$
and at $|R|=20K/h$.
 The evolution of $\epsilon$ can be quite different from
$\epsilon_r$ if we use  the temperature cycle proposed in
ref.\cite{VincentPRL}. After a cooling at $R=-20K/h$ from
$T_{max}$ to $T_{stop}=374K$ the sample is maintained at
$T_{stop}$ for $10h$. After this time interval the sample is
cooled again, at the same $R$, down to $T_{min}$. Once the sample
temperature reaches $T_{min}$ the sample is heated again at
$R=20K/h$ up to $T_{max}$. The dependence of $\epsilon$ as a
function of $T$, obtained when the sample is submitted to this
temperature cycle with the cooling stop at $T_{stop}$, is called
the memory curve $\epsilon_m$. In fig.\ref{fig:tempcyc2}(a),
$\epsilon_m$ (solid line), measured at $\nu=0.1 Hz$, is plotted as
a function of $T$. The dashed line corresponds to the reference
curve of fig.\ref{fig:tempcyc1}(a).
 We notice that
$\epsilon_m$ relaxes downwards when cooling is stopped at
$T_{stop}$: this corresponds to the vertical line  in
fig.\ref{fig:tempcyc2}(a) where $\epsilon_m$ departs from
$\epsilon_r$. When cooling is resumed $\epsilon_m$ merges into
$\epsilon_r$ for $T <340K$. The aging at $T_{stop}$ has not
influenced the result at low temperature. This effect has been
called {\it rejuvenation} in recent papers \cite{Dupuis,Bouchaud}.

\begin{figure}
    \begin{center}
    {\psfrag{xl}[ct][ct]{\small Temperature $T$ (K)}
            \psfrag{yl}[Bc][Bc]{\small $\epsilon_{mc} - \epsilon_{rc}$}
            \includegraphics[width=6cm]{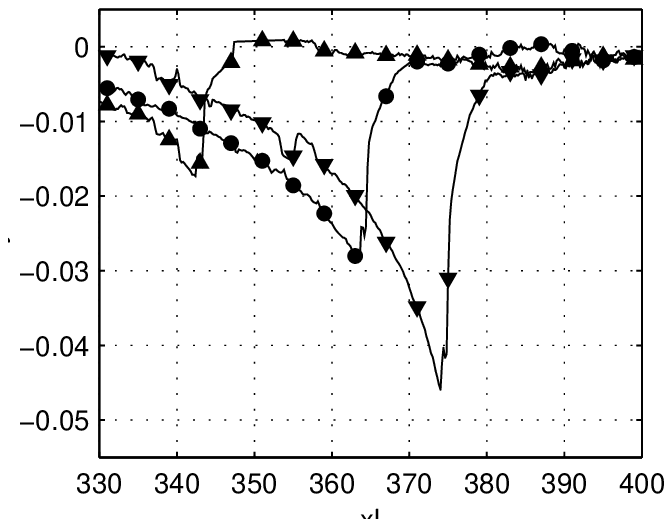}}
        \hspace{4mm}
        {
            \psfrag{xl}[ct][ct]{\small Temperature $T$ (K)}
            \psfrag{yl}[Bc][Bc]{\small $\epsilon_{mh} - \epsilon_{rh}$}
            \includegraphics[width=6cm]{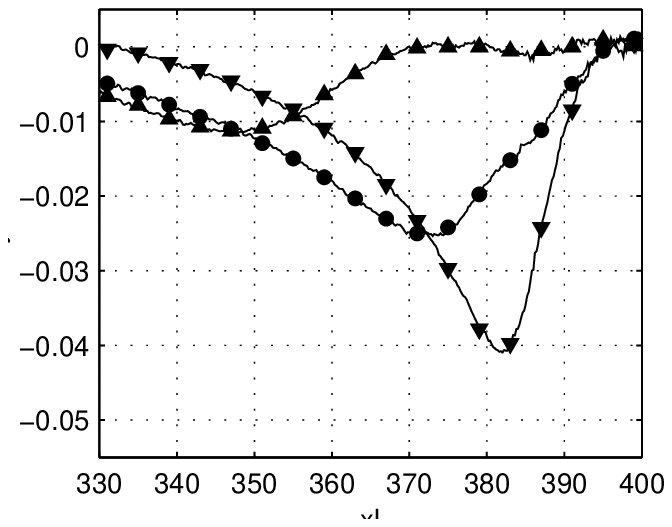}
        }
    \end{center}
\caption{Dependence on $T_{stop}$. Difference between $\epsilon_r$
and $\epsilon_m$ for 3 different cooling stops measured at
$\nu=0.1Hz$ for $|R|=10K/h$ and $t_{stop}=10h$. (a) Writing memory
(cooling): $\epsilon_{mc} - \epsilon_{rc}$ with $T_{stop} = 344 K$
($\blacktriangle$), $364 K$ ($\bullet$) and $374 K$
($\blacktriangledown$). (b) Reading memory (heating):
$\epsilon_{mh} - \epsilon_{rh}$ (same symbols as in (a) ).
} \label{fig:Tstopdependance}
\end{figure}

During the heating period the system reminds the aging at
$T_{stop}$ (cooling stop) and for $340K<T<395K$ the evolution of
$\epsilon_m$ is quite different from $\epsilon_r$. In order to
clearly see this effect we divide $\epsilon_m$ in the cooling part
$\epsilon_{mc}$ and the heating part $\epsilon_{mh}$. In
fig.\ref{fig:tempcyc2}(b) we plot the difference between
$\epsilon_m$ and $\epsilon_r$. Filled downwards arrows correspond
to cooling ($\epsilon_{mc}-\epsilon_{rc}$) and empty upward arrows
to heating ($\epsilon_{mh}-\epsilon_{rh}$). The difference between
the evolutions corresponding to different cooling procedures is
now quite clear. The system reminds its previous aging history
when it is reheated from $T_{min}$. The amplitude of the memory
corresponds well to the amplitude of the aging at $T_{stop}$ but
the temperature of the maximum is shifted a few degrees above
$T_{stop}$. On fig.\ref{fig:Tstopdependance} we show that this
temperature shift is independent of $T_{stop}$ for temperatures
where aging can be measured in a reasonable time (above $335K$).
In contrast the amplitude of the downward relaxation at $T_{stop}$
is a decreasing function of $T_{stop}$, as expected from the first
section measurements: no aging can be measured at $T_{stop}<335K$
(see fig.\ref{fig:alpha}(a)).

\begin{figure}
\begin{center}
\epsfig{figure=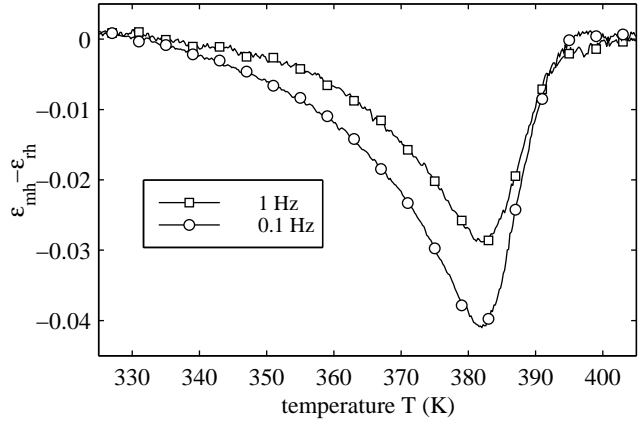}
\end{center}
\caption{Dependence on frequency. Reading the memory (difference between
heating curves $\epsilon_{mh} -\epsilon_{rh}$) after a 10h stop at
$T_{stop}=374K$ during cooling. The same rate of $10K/h$ is used but the
measurement is done at different frequencies: $\nu=1Hz$ ($\square$) and
$\nu=0.1Hz$ ($\circ$).}
\label{fig:frequ}
\end{figure}

This memory effect seems to be permanent because it does not depend on the
waiting time at $T_{min}$. Indeed we performed several experiments in which we
waited till $24h$ at $T_{min}$, before restarting heating, without noticing
any change in the heating cycle. In contrast the amplitude and the position of
the memory effect depend on $R$ and on the measuring frequency. As an example
of frequency dependence, we compare in fig.\ref{fig:frequ} two measurements
done with the same $|R|=10K/h$ and waiting time $t_{stop}=10h$ but for two
different frequencies: $\nu=0.1Hz$ and $\nu=1Hz$. Again, as expected from the
first section measurements, the higher is the frequency, the smaller is the
downward relaxation, so the smaller is the memory effect. The positions of the
maxima are at the same temperature.

\begin{figure}
\begin{center}
\epsfig{figure=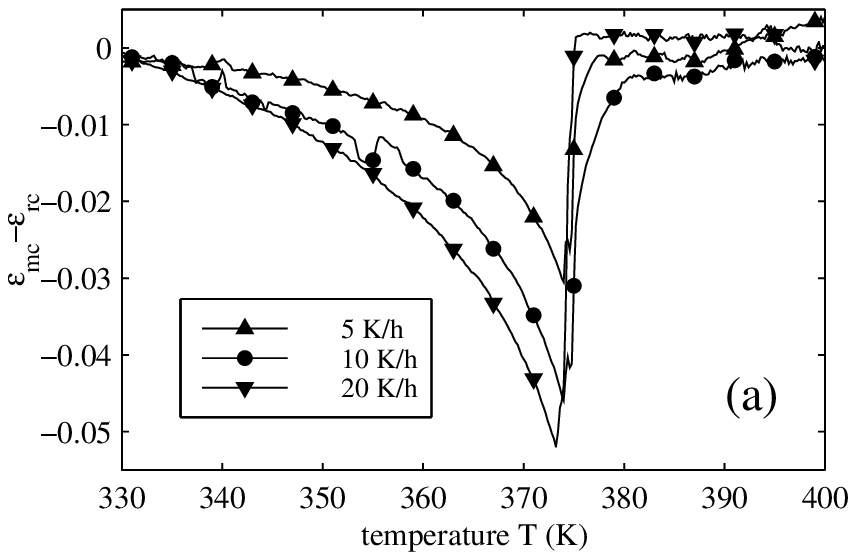} \epsfig{figure=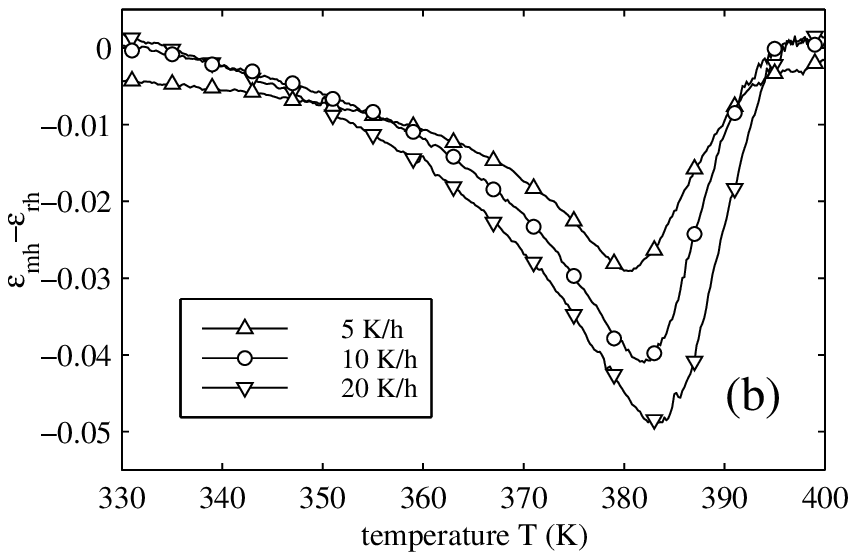}
\end{center}
\caption{Dependence on the cooling and heating rate. Difference between
$\epsilon_r$ and $\epsilon_m$ (aging at $T_{stop}=374K$ for $10h$) measured at
$\nu=0.1Hz$ for 3 $|R|$. (a) Writing memory (cooling): $\epsilon_{mc} -
\epsilon_{rc}$ at $5 K/h$ ($\blacktriangle$), $10 K/h$ ($\bullet$) and $20
K/h$ ($\blacktriangledown$). (b) Reading memory (heating): $\epsilon_{mh} -
\epsilon_{rh}$ at $5 K/h$ ($\vartriangle$), $10 K/h$ ($\circ$) and $20 K/h$
($\triangledown$).
 }
\label{fig:rate}
\end{figure}

As an example of rate dependence, at $\nu=0.1Hz$ and $t_{stop}=10h$, we plot
in fig.\ref{fig:rate} the difference $\epsilon_m-\epsilon_r$ as a function of
$T$ for three different rates. The faster is the rate, the larger is the
downward relaxation of the dielectric constant during the cooling stop. As the
amplitude of the memory effect is equal to that of the relaxation, we just
expect the memory effect to increase with $|R|$. But as we can see on
fig.\ref{fig:rate}, the temperature positions of the maxima are rate dependent
too: the larger is $|R|$, the farther the temperature of the maximum is
shifted above the aging temperature $T_{stop}$. The cooling rate is not the
only control parameter of the memory effect, the heating rate is relevant too.

\section{Advanced memory experiments}

In this section we apply to the PMMA sample more complicated
temperature histories (inspired from spins glasses experiments
\cite{VincentPRL,Jonason2000}): What happens if we try to read a
memory twice ? If we make two cooling stops ?

\subsection{Deleting memory}

\begin{figure}
    \begin{center}
         \psfrag{2}[bl][bl]{ }
         \psfrag{(a)}[bl][bl]{(a)}
         \psfrag{(b)}[bl][bl]{(b) }
         \psfrag{I}[cc][cc]{\colorbox{white}{{\small I}}}
         \psfrag{II}[cc][cc]{\colorbox{white}{{\small II}}}
         \psfrag{i}[cc][cc]{\colorbox{white}{{\tiny I}}}
         \psfrag{ii}[cc][cc]{\colorbox{white}{{\tiny II}}}
         \psfrag{Td}[cc][cc]{\colorbox{white}{\small $T_i$}}
         \psfrag{Tstop}[cc][cc]{\colorbox{white}{\small $T_{stop}$}}
         \psfrag{Tde}[cc][cB]{\colorbox{white}{\tiny $T_i$}}
         \psfrag{Tstope}[cc][cc]{\colorbox{white}{\tiny $T_{stop}\!\!$}}
         \psfrag{Tge}[cc][cc]{\colorbox{white}{\tiny $T_{g}$}}
         \psfrag{xl}[ct][cc]{\small Temperature $T$ (K)}
         \psfrag{yl}[Bc][Bc]{\small $\epsilon_{mh} - \epsilon_{rh}$}
         \psfrag{xle}[ct][ct]{\colorbox{white}{\tiny Time $t$ (h)}}
         \psfrag{yle}[Bc][Bc]{\colorbox{white}{\tiny Temperature $T$ (K)}}
         \includegraphics[width=11cm]{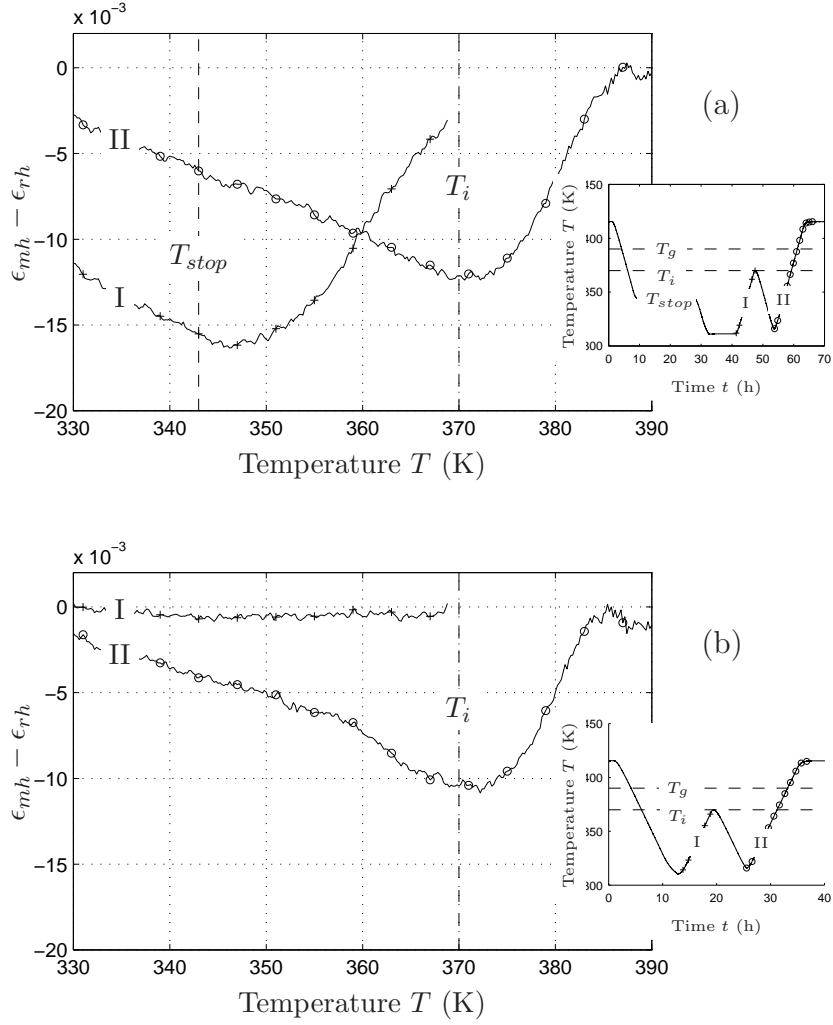}
    \end{center}
\caption{Deleting a memory by reading it. (a) Difference between
heating curves $\epsilon_{rh}$ and $\epsilon_{mh}$ after the
temperature history shown in the inset: after a classic reading
the memory of a $20h$ stop at $T_{stop}=345K$ with $|R|=20K/h$ and
$\nu=0.1Hz$ ($+$), the sample is cooled again when the temperature
reaches $T_i<T_g$. The second reading ($\circ$) is very different
since no tracks of $T_{stop}$ is found but a sort of memory of the
inversion temperature $T_i$. This can be checked on (b), where we
only test the memory of the inversion. The difference between the
second heating curves ($\circ$) $\epsilon_{mh} - \epsilon_{rh}$ is
exactly the same for both temperature histories. Heating the
sample up to $T_i$ reinitializes the lower temperature behavior,
even though $T_i < T_g$.} \label{fig:delete}
\end{figure}

In this experiment, we show that a memory can be read only one
single time: reading a memory effect also deletes it. This
experiment is inspired from similar ones done in spin glasses
\cite{Jonason2000}. First we follow the classic procedure
described in the previous section: during the cooling ramp at
$R=-20 K/h$, a temporary stop is done for $t_{stop}=20h$ at
$T_{stop}=345K$, and then heating the sample from $T_{min}$ with
the same $|R|$. In fig.\ref{fig:delete}(a)  we plot the difference
between the heating branches of $\epsilon_m$ and $\epsilon_r$
measured at $\nu=0.1Hz$. The departure from the reference curve
above $T_{stop}$ when lowering temperature is due to a smooth
cutting of cooling, but  this imperfection has no detectable
effect on the heating curve. We can follow the memory of the
cooling stop at $T_{stop}$, and when the memory curve almost
merges the reference one, for $T_i=368K$, we quickly stop heating
and resume cooling at $R=-20K/h$. Notice  that this inversion
temperature $T_i$ is smaller than $T_g$, which means that the
sample has not been reinitialized. When $T_{min}$ is reached
again, we make a classic heating at $R=+20K/h$ (see the
temperature history of the sample in the inset of
fig.\ref{fig:delete}(a)).

This second heating curve is really different from the first one:
there are no tracks of the cooling stop at $T_{stop}=345K$, but
something like the memory of a stop around $370K$. What we see now
is in fact only the memory of the stop at $T_i$: as the sample
stays a few minutes around $T_i$ (the time needed to inverse $R$
to $-R$), it ages a little at $T_i$ and we find a memory of this
event. This can be checked on fig.\ref{fig:delete}(b), where we
only show the memory of the inversion at the same temperature
$T_i$, without the first memory. The curves corresponding to the
second heating are exactly the same for the two experiments,
showing that the first heating deletes all information about
temperatures lower than $T_i$, even though $T_i<T_g$.

\subsection{Double memory}

\begin{figure}
    \begin{center}
         \psfrag{2}[bl][bl]{ }
         \psfrag{xl}[ct][ct]{\small Temperature $T$ (K)}
         \psfrag{yl}[Bc][Bc]{\small $\epsilon_{mc} - \epsilon_{rc}$}
         \psfrag{xle}[ct][ct]{ {\tiny Time $t$ (h)}}
         \psfrag{yle}[Bc][Bc]{\tiny Temperature $T$ (K)}
         \includegraphics[width=11cm]{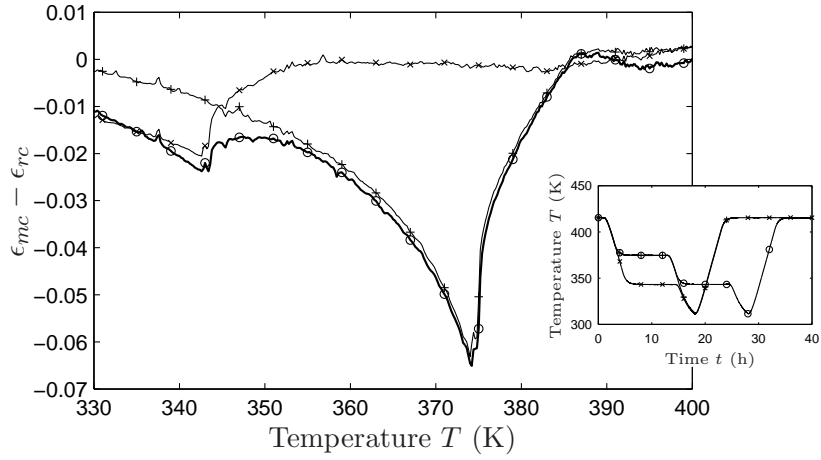}
    \end{center}
\caption{Double memory recording. Difference between cooling
curves $\epsilon_{mc}$ and $\epsilon_{rc}$ during the temperature
history shown in the inset, for $|R|=20K/h$, and $\nu=0.1Hz$: one
$10h$ stop at $T_{stop1}=375K$ ($+$), one $10h$ stop at
$T_{stop2}=345K$ (x), and two $10h$ stops at $T_{stop1}=375K$ and
$T_{stop2}=345K$ ($\circ$). The low temperature state is
independent of high temperature history: aging during $10h$ at
$T_{stop2}$ produces the same relaxation, whatever happened before
at higher temperatures.} \label{fig:doublerecord}
\end{figure}

More insight on the properties of the memory effects can be
obtained by submitting the sample to a more complex cooling
procedure consisting of two cooling stops. This procedure, which
has been called the double memory effect, has been carried out
successfully in spin glasses \cite{VincentPRL} where it has been
observed that if two stops are done during cooling, the heating
curve will present a memory effect for both stops. The difficulty
that arise when trying to reproduce this experiment in PMMA is the
narrowness of the aging range: when cooling the sample under
$T_{stop}$, $\epsilon_m$ rejoins the reference curve only for
temperatures where aging almost vanishes. It is therefore
difficult to record two well distinct cooling stops. This is
illustrated in fig.\ref{fig:doublerecord}, where $T_{stop1}=375K$
and $T_{stop2}=345K$: $\epsilon_m$ has not completely merged into
$\epsilon_r$ when we stop cooling for the second time.

The double memory experiments allows us to point out  another
important property of PMMA aging: when temperature is lowered
after the first stop, the system not only recover the same value
of $\epsilon$, but also the same aging properties. Indeed, a $10
h$ stop at $345K$ produces the same downward relaxation of
$\epsilon$, whatever the previous history is. The low temperature
state is thus completely uninfluenced by the high temperature
history.

\begin{figure}
    \begin{center}
         \psfrag{xl}[ct][ct]{\small Temperature $T$ (K)}
         \psfrag{yl}[Bc][Bc]{\small $\epsilon_{mh} - \epsilon_{rh}$}
         \includegraphics{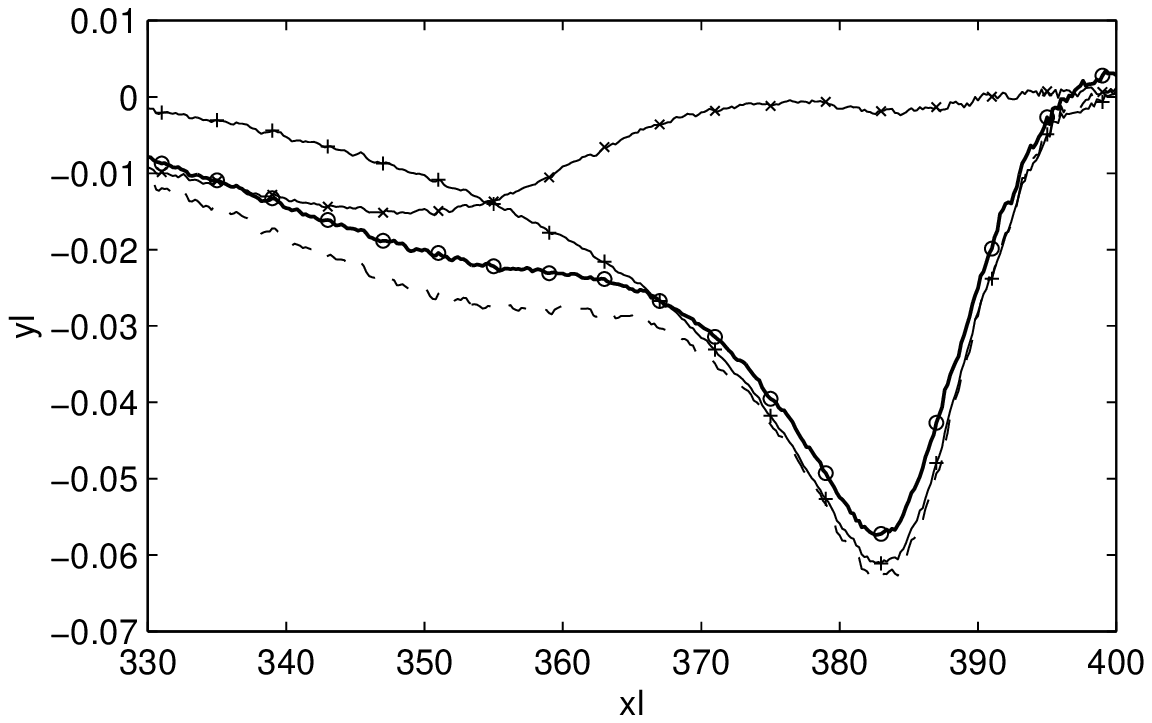}
    \end{center}

\caption{Double memory reading. Difference between heating curves
$\epsilon_{mh}$ and $\epsilon_{rh}$ corresponding to the cooling
curves of fig.\ref{fig:doublerecord}: reading of a single stop at
$T_{stop1}=375K$ ($+$) and $T_{stop2}=345K$ (x), reading of a
double stop at $T_{stop1}$ and $T_{stop2}$ ($\circ$). The dashed
line is the sum of the two single stop curves, it is a very good
approximation of the double memory reading. Memory effects thus
appear only additive in this case, where $T_{stop1}$ and
$T_{stop2}$ are sufficiently far from each other.}
\label{fig:doubleread}
\end{figure}

If we now heat the sample after the two cooling stops, we obtain
the bolt curve of fig.\ref{fig:doubleread}. If the memory of
$T_{stop1}$ is obvious, the lower temperature stop at $T_{stop2}$
must be hidden in the first part of the curve. In order to check
the presence of the memory effect of the second stop, let us
suppose that this effect is just additive, that is the memory of a
double stop is just the sum of the memory of both individual stops
if their temperature is sufficiently different. So we plot with a
dashed line in fig.\ref{fig:doubleread} the sum of the single
memories of $T_{stop1}$ and $T_{stop2}$. Within errors bars, the
two curves are the same, so we conclude that even if the memory
effects overlap in the narrow aging range of PMMA, double memory
experiments also work in this polymer.

\section{Discussion and conclusions}

We have applied to a PMMA sample  the same temperature cycle used
to study memory and rejuvenation effects in various kinds of spin
glasses and in orientational glasses. Our dielectric measurements
clearly show the presence of rejuvenation and memory in PMMA. As
we already mentioned in the introduction the observation of a
memory of the thermal history in PMMA is not new. However this
memory effect has been obtained with a thermal protocol which is
quite different from the one described in this paper. Thus the
application to a polymer glass of the same procedure used in OG
and SG allow us to address the question of the universality of the
memory and rejuvenation phenomena in different materials. This is
a very important point in to understand whether the same
theoretical approach can be used to describe aging phenomena in
different materials.

Let us summarize the main results of these low frequency
dielectric measurements on PMMA:


\begin{itemize}
  \item[a)]
The reference curve $\epsilon_r$, obtained at constant cooling and
heating rate $|R|$ is hysteretic. This hysteresis is maximum a few
degrees above $T_g$.
  \item[b)]
The hysteresis of $\epsilon_r$ increases with $|R|$.
  \item[c)]
Writing memory: a cooling stop produces a downward relaxation of
$\epsilon_m$. The amplitude of this downward relaxation depends on
$T_{stop}$ and it decreases for decreasing $T_{stop}$. It almost
disappears for $T_{stop}<335K$.
  \item[d)]
When cooling is resumed $\epsilon$ goes back to the cooling
branch of the reference curve. This suggests that the low
temperature state is independent on the cooling history.
  \item[e)]
Reading memory: upon reheating $\epsilon_m$ reminds the aging
history and the cooling stop ({\it Memory}). The maximum of the
memory effect is obtained a few degrees above $T_{stop}$.
  \item[f)]
The memory effect does not depend on the waiting time at low
temperature but it depends both on the cooling and heating rates.
The memory effect increases with $|R|$.
  \item[g)] Double memory effects are observed with more difficulty in PMMA
than in SG. The difference comes from the fact that aging effects
are reduced when temperature is lowered. However  a careful
analysis of experimental data allows us to show the existence of
double memory effects in PMMA.
\item[h)] The memory effect is deleted by a reading, even if the
temperature remains smaller than $T_g$.
\end{itemize}

Analogies between point a-b)  for the hysteresis and point e-f)
for the rate dependence of the memory effect leads to a new
interpretation of hysteresis, which can be seen as the memory of
aging at a temperature $T_{stop}\approx T_g$. Indeed in a free
energy landscape model, a  sample, which is cooled  just above
$T_g$, is in its equilibrium phase, that is in a favorable
configuration at this temperature. If this configuration is not
strongly modified by aging at lower temperatures then, when
heating back to $T_g$, the system reminds this favorable state,
just as it does in the memory effect.

It is interesting to discuss the analogies and the differences
between this experiment and similar ones performed on SG
\cite{VincentPRL,Jonsson1,Dupuis} and on OG \cite{Doussineau}. It
turns out that, neglecting the hysteresis of the reference curve
of PMMA and of OG, the behavior of these materials is quite
similar to that of SG. A strong rate dependence has been observed
in Ising spin glasses too \cite{Dupuis}. During the heating period
PMMA, SG and OG remind their aging history, although the precise
way, in which history is remembered, is material dependent.
Furthermore in these materials the low temperature state is
independent on the cooling history. One can estimate the
temperature range $\delta T$ where the material response is
different from that of the reference curve because of the cooling
stop. It turns out that the ratio $\delta T/ T_G $ is roughly the
same in PMMA, in SG and in OG, specifically $\delta T /T_G \simeq
0.2$. The important difference between SG and PMMA is in the
dependence on $T_{stop}$ of the amplitude of the downward
relaxation: it is strong  in PMMA and weak in SG.

Our results seem to indicate that memory and rejuvenation
phenomena  in the aging process may be described by models based
on a hierarchical free energy landscape, whose barriers grow when
temperature is lowered \cite{Vincent,VincentPRL}. However the
dependence of the memory effect on $|R|$ and the independence on
the waiting time at $T_{min}$ mean that, at least for PMMA, the
free energy landscape has to depend not only on temperature but
also on $|R|$. Many models \cite{Vincent,Mezard,Fisher,Bray} and
numerical simulations \cite{Marinari,Barrat} do not take into
account this dependence because they consider just a static
temperature after a quench. In contrast point f) indicates that
the cooling history is relevant too.

These difficulties can be avoided if one considers models based on
a slow domain growth and domain walls reconformations in the
pinning field created by disorder (see for example
ref.\cite{Bouchaud1} and references therein). These models imply
the existence of a hierarchy of length scales $l$ with a
characteristic time $\tau \propto l^z$. Recent numerical
simulations \cite{Berthier} show that this is an important
ingredient in order to have memory/rejuvenation effects. However
one of the drawbacks of such  a model is that memory is recovered
only if the time spent at low temperature is short
enough\cite{Bouchaud1}. This effect does not seem to be true for
PMMA, at least on reasonable laboratory time scales.

As a conclusion the "memory" and rejuvenation effects seem to be
two universal features of aging whereas the hysteresis is present
in PMMA and in OG but not in all kinds of spin glasses. It would
be interesting to know whether  these effects are observed in
other polymers and in supercooled liquids, and whether the
hysteresis interpretation in terms of a memory effect could hold
for other materials. These measurements show that many features of
aging  seem to be "universal" in  several materials and that
models based on spin glasses may be  useful to describe aging in
polymeric materials.

\bigskip

{\bf Acknowledgements}

 We acknowledge useful
discussions with J. P. Bouchaud, J. Kurchan, M. M\'ezard, G. Vigier and
E. Vincent. We thank P. Metz and L. Renaudin for technical
support. This work has been partially supported by the R\'egion
Rh\^one-Alpes contract ``Programme Th\'ematique: Vieillissement
des mat\'eriaux amorphes'' .

\newpage


\newpage


\begin{thebibliography}{}


\bibitem{Struick} L.C. Struick, {\it Physical aging
in amorphous polymers and other materials} (Elsevier, Amsterdam,
1978).

\bibitem{book}
 {\it Spin Glasses and Random Fields}, edit by A. P. Young, Series
 on Directions in Condensed Matter Physics Vol.12 ( World Scientific, Singapore
 1998).

\bibitem{Vincent} M. Lederman, R. Orbach, J.M. Hammann, M. Ocio,
 E. Vincent, Phys. Rev. B, {\bf 44}, 7403 (1991); E. Vincent, J. P. Bouchaud,
 J. Hammann, F. Lefloch, Phil. Mag. B, {\bf 71}, 489 (1995).
 ; C. Djuberg,
 K. Jonason, P. Nordblad, cond-mat/9810314.

\bibitem{Alberici} F. Alberici, P. Doussineau, A. Levelut
 Europhysics Lett. {\bf 39}, 329 (1997).

\bibitem{Nagel} R. L. Leheny, S. R. Nagel, Phys. Rev.B {\bf 57}, 5154
(1998).

\bibitem{VincentPRL} K.Jonason, E. Vincent, J. Hamman, J. P.
Bouchaud, P. Nordblad, Phys. Rev. Lett. 81, 3243 (1998).


\bibitem{Jonsson1}T. Jonsson, K. Jonason, P. Nordblad,
Phys. Rev. B 59, 9402 (1999); T. Jonsson, K. Jonason, P. Jonsson,
P. Nordblad, Phys Rev. B 59,8770 (1999).

\bibitem{Doussineau} P. Doussineau, T. Lacerda-Aroso, A. Levelut,
Europhys. Lett., 46, 401 (1999).

\bibitem{vigier} E. Muzeau, G. Vigier, R. Vassoille and J. Perez,
 Polymer,  36,  (1995), 611.

\bibitem{Bellon} L. Bellon, S. Ciliberto, C. Laroche,
cond-mat/9905160.

\bibitem{Europhys}  L. Bellon, C. Laroche, S. Ciliberto, Europhys.
Lett. {\bf 51}, 551 (2000)

\bibitem{Jonason2000} K. Jonason, P. Nordblad, E. Vincent, J. Hammann
and J.-P. Bouchaud, Eur. Phys. J. B. 13, 99 (2000)

\bibitem{Dupuis} V. Dupuis, E. Vincent, J.P. Bouchaud, J. Hammann,
A. Ito, H. Aruga Katori, Aging, rejuvenation and memory effects in
Ising and Heisenberg spin glasses. cond-mat/0104399

\bibitem{Bouchaud} J. P. Bouchaud, L.F. Cugliandolo, J. Kurchan,
M. M\'ezard, in {\it Spin Glasses and Random Fields}, \cite{book},
and references therein.

\bibitem{Mezard} M.M\'ezard, G. Parisi, M. A. Virasoro, in Spin Glasses
 Theory and Beyond, World Scientific Lecture Notes in Physics
 Vol.9 ( World Scientific, Singapore 1987).

\bibitem{Bouchaud1} J. P. Bouchaud," Aging in glassy systems: new
experiments, simple models and open questions ", cond-mat/9910387
and 'Soft and Fragile Matter: Nonequilibrium Dynamics,
Metastability and Flow', M. E. Cates and M. R. Evans, Eds., IOP
Publishing (Bristol and Philadelphia) 2000, pp 285-304

\bibitem{Kurchan} L. F. Cugliandolo, J. Kurchan, cond-mat/9812229
(1998)

\bibitem{Kovacs} A. Kovacs, J. Polym. Sci. 30 (1958) 131.



\bibitem{bookp} N. G. McCrum, B. E. Read, G. Williams
{\it Anelastic and Dielectric Effects in Polymeric Solids},(Dover
1991)


\bibitem{Fisher} D. S. Fisher, D. A. Huse, Phys. Rev. Lett. 56,
 1601, (1987).

\bibitem{Bray} A. J. Bray, M. A. Moore, Phys. Rev. Lett. 58, 57
 (1987).

\bibitem{Marinari} E. Marinari, G. Parisi, J.J. Ruiz-Lorenzo, F.
Ritort Phys. Rev. Lett. 76, 843 (1996); E. Marinari, G. Parisi
J.J. Ruiz-Lorenzo, in {\it Spin Glasses and Random Fields}
pp.59-98; E. Marinari, G. Parisi, J. J. Ruiz-Lorenzo Phys. Rev. B.
58, 14852 (1998).

\bibitem{Barrat} W. Kob, J.L. Barrat, Phys. Rev. Lett. 78, 4581
(1997).

\bibitem{Berthier} L. Berthier, P. C. Holdsworth, cond-mat/0109169v1

\end{thebibliography}
\end{document}